\documentclass[useAMS,usenatbib,epsfig]{mn2e}
\def\gtrsim{\mathrel{\hbox{\rlap{\hbox{\lower4pt\hbox{$\sim$}}}\hbox{$>$}}}}
\usepackage[latin1]{inputenc} 

\def\lesssim{\mathrel{\hbox{\rlap{\hbox{\lower4pt\hbox{$\sim$}}}\hbox{$<$}}}}

\newif\ifAMStwofonts
\AMStwofontstrue
\usepackage{amsmath,amsfonts,epsfig,natbib}

\def\reff@jnl#1{{\rm#1\/}}
\def\aj{\reff@jnl{AJ}}                  
\def\araa{\reff@jnl{ARA\&A}}            
\def\apj{\reff@jnl{ApJ}}                        
\def\apjl{\reff@jnl{ApJ}}               
\def\apjs{\reff@jnl{ApJS}}              
\def\ao{\reff@jnl{Appl.Optics}}         
\def\apss{\reff@jnl{Ap\&SS}}            
\def\aap{\reff@jnl{A\&A}}               
\def\aapr{\reff@jnl{A\&A~Rev.}}         
\def\aaps{\reff@jnl{A\&AS}}             
\def\azh{\reff@jnl{AZh}}                        
\def\baas{\reff@jnl{BAAS}}              
\def\gca{\reff@jnl{GeCoA}}              
\def\jrasc{\reff@jnl{JRASC}}            
\def\memras{\reff@jnl{MmRAS}}           
\def\mnras{\reff@jnl{MNRAS}}            
\def\physrep{\reff@jnl{Phys.~Rep.}}     
\def\pra{\reff@jnl{Phys.Rev.A}}         
\def\prb{\reff@jnl{Phys.Rev.B}}         
\def\prc{\reff@jnl{Phys.Rev.C}}         
\def\prd{\reff@jnl{Phys.Rev.D}}         
\def\prl{\reff@jnl{Phys.Rev.Lett}}      
\def\pasp{\reff@jnl{PASP}}              
\def\pasj{\reff@jnl{PASJ}}              
\def\qjras{\reff@jnl{QJRAS}}            
\def\skytel{\reff@jnl{S\&T}}            
\def\solphys{\reff@jnl{Solar~Phys.}}    
\def\sovast{\reff@jnl{Soviet~Ast.}}     
\def\ssr{\reff@jnl{Space~Sci.Rev.}}     
\def\zap{\reff@jnl{ZAp}}                        
\def\nat{\reff@jnl{Nature}}             

\title[The distribution of galaxies in the VSA cold spot]
{A study of the galaxy redshift distribution toward the cosmic microwave background cold spot in 
the Corona Borealis supercluster}
\author[R. G\'enova-Santos et al.] {Ricardo G\'enova-Santos,$^{1,2}\thanks{E-mail: rgs@iac.es}$ Carmen Pilar Padilla-Torres,$^{1,2}$  
\newauthor Jos\'e Alberto Rubi\~no Mart\'{\i}n,$^{1,2}$
Carlos M. Guti\'errez$^{1,2}$ and Rafael Rebolo$^{1,2,3}$\\ 
$^1$ Instituto de Astrofis\'{\i}ca de Canarias, 38200 La Laguna, Tenerife, Canary Islands, Spain \\
$^2$ Universidad de La Laguna, Spain\\
$^3$ Consejo Superior de Investigaciones Cient\'{\i}ficas, Spain 
}  
\date{Accepted Received In original form}
\pagerange{\pageref{firstpage}--\pageref{lastpage}}
\pubyear{2009}

\begin{document}

\label{firstpage}
\maketitle

\begin{abstract}
We present a study of the spatial and redshift distributions of Sloan Digital Sky Survey (SDSS) galaxies  
toward the position of CrB-H, a very deep and extended decrement in the Cosmic Microwave Background 
(CMB), located within the Corona Borealis supercluster (CrB-SC). It 
was found in a survey with the Very Small Array (VSA) interferometer at 33~GHz, with a peak negative 
brightness temperature of $-230~\mu$K, and deviates 4.4$\sigma$ from the Gaussian CMB 
(G\'enova-Santos et al.).   
Observations with the Millimeter and Infrared Testa Grigia Observatory (MITO) suggested 
that 25$^{+21}_{-18}\%$ of this decrement may be caused by the thermal Sunyaev-Zel'dovich (tSZ) effect (Battistelli et al.). 
Here we investigate whether the galaxy distribution 
could be tracing either a previously unnoticed galaxy cluster or a Warm/Hot Intergalactic Medium (WHIM) 
filament that could build up this tSZ effect. 

We find that the projected density of galaxies outside Abell clusters and with redshifts $0.05<z<0.12$ at the position 
of CrB-H is the highest in the area encompassed by the CrB-SC. Most of these galaxies are located around redshifts 
$z=0.07$ and $z=0.11$, but no clear connection in the form of a filamentary structure is appreciable in between.
While the galaxy distribution at $z=0.07$ is sparse, we find evidence at $z=0.11$ of a galaxy group or a low-mass 
galaxy cluster. We estimate that this structure could produce a tSZ effect of $\approx -18~\mu$K. 
The remaining VSA signal of $\approx -212~\mu$K is still a significant  4.1$\sigma$ deviation from the Gaussian CMB. 
However, the MITO error bar allows for a larger tSZ effect, which could 
be produced by galaxy clusters or superclusters beyond the sensitivity of the SDSS. Contributions from 
other possible secondary anisotropies associated with these structures are also discussed.

Subject headings: galaxies: clusters: individual
galaxies: distances and redshifts large-scale structure of universe surveys 
\end {abstract}

\section{Introduction}

In a survey in the Corona Borealis supercluster (CrB-SC) performed with the Very Small Array (VSA) 
interferometer at 33~GHz and with an angular resolution of FWHM$\approx 11$~arcmin, \citet{genova_05} 
discovered a remarkably large and deep 
decrement (the so-called CrB-H decrement) in the Cosmic Microwave Background (CMB), with a minimum brightness 
temperature of $-230\pm 23~\mu$K, angular size of $\approx 30$~arcmin and coordinates 
RA=15$^{\rm h}$22$^{\rm m}$11.47$^{\rm s}$, Dec.=28$^\circ$54'06.2''. Subsequent observations with the VSA at the same
frequency but with a finer angular resolution of FWHM$\approx 7$~arcmin \citep{genova_08}, and with the 
Millimeter and Infrared Testa Grigia Observatory (MITO) 
telescope at 143, 214 and 272~GHz and FWHM$\approx 16$~arcmin
\citep{battistelli_06}, confirmed the presence of this feature. Statistical analyses based on Monte Carlo simulations 
indicated that this decrement is a significant deviation from the Gaussian CMB at the level of 4.4$\sigma$.
An independent analysis based on smooth goodness-of-fit tests yielded a deviation with respect to Gaussianity of 99.8~per
cent \citep{rubino_06}. A possible explanation for at least part of this decrement 
is the Sunyaev-Zel'dovich (SZ; Sunyaev \& Zel'dovich 1972) effect, a secondary anisotropy in the CMB
that generates temperature decrements at the VSA frequency. This effect is produced by the inverse Compton scattering of the CMB
photons by hot electrons, and is split into two components: the thermal SZ (tSZ) effect, produced by the thermal motion of the
electrons, and the kinematic SZ (kSZ) effect, due to the peculiar motion of the structure containing the scattering electrons.
By considering the tSZ characteristic spectral dependency, together with the flat-spectrum of the primordial CMB,
\citet{battistelli_06} estimated from the MITO observations that 25$^{+21}_{-18}$\% of the total observed decrement is compatible with a tSZ
component. The tSZ occurs in the cores of rich galaxy
clusters, where it has been extensively detected and studied (e.g. Lancaster et al. 2005). However, there are no Abell clusters at this position, 
nor significant X-ray 
emission in the ROSAT data \citep{genova_05}, as it would be expected in the case of the presence of a nearby rich SZ cluster. 
For this reason, if the SZ effect is the responsible, at least partially, for the detected decrement, then it would have to 
arise from either a distant unknown cluster or a less hot and dense gas distribution with a weak X-ray emission. In fact, less hot and dense
structures like superclusters of galaxies are predicted to build up detectable SZ signals thanks to the long paths of the CMB
photons across them \citep{birkinshaw_99}. Therefore, the CrB-H decrement could 
be indicative of the presence of a warm/hot gas distribution in the intercluster medium with a high elongation toward the
line of sight. A confirmation of this hypothesis
is in order, as structures like this one could provide the location for a significant fraction of the baryon content in the
Local Universe. In fact, about half of the total baryonic matter at $z=0$ remains undetected by the usual observational
methods (e.g. Fukugita et al. 1998). According to cosmological simulations \citep{cen_99,dave_01}, most of these 
``missing baryons'' (30-50\%) could be located in the so-called ``warm/hot intergalactic medium'' (WHIM)
as a low-density plasma forming large-scale filamentary structures connecting galaxy clusters.

However, the temperature and baryon density of the hypothetical WHIM structure required to produce a tSZ signal like the detected
decrement are in tension with the simulations, which predict baryon overdensities in the range 
$\delta\rho_{\rm B}/\langle\rho_{\rm B}\rangle\sim 10-30$ and temperatures $10^5\le T \le 10^7$~K. On the other hand, the 
possibility of a farther background galaxy cluster is difficult to reconcile with the large angular extension 
($\sim 25$~arcmin) of the decrement. For this reason, the idea of a combination between primordial CMB and a tSZ 
effect, which 
was supported by the MITO observations, seems the most plausible hypothesis to explain the observations without violating the 
results from the simulations. However, the uncertainty interval of MITO tSZ relative contribution 
would leave the primordial CMB component as a significant $2.3-4.1\sigma$ deviation from the concordance model.
Recently, \citet{flores_09} analyzed regions similar to CrB-SC in the {\it MareNostrum Universe} \footnote{\tt
http://astro.ft.uam.es/$\sim$marenostrum/} simulation, and concluded that low dense WHIM regions can 
produce
at most $\approx -10~\mu$K SZ decrements, and that the CrB-H feature can most likely be generated by galaxy clusters.
A detailed optical study of this region can help to discriminate between the WHIM and cluster 
hypotheses, as it could reveal the presence of either an sparse distribution of galaxies along the line of sight tracing a
filamentary structure or of a farther galaxy cluster. 
\citet{gal_03} and \citet{koester_07} 
tentatively identified galaxy clusters near the coordinates of the decrement at $z\approx 0.07$ and $z\approx 0.11$, by using 
galaxy catalogues from the Second Palomar Observatory Sky Survey and by applying the maxBCG red-sequence method to the Sloan 
Digital Sky Survey (SDSS), respectively.
\citet{padilla_09} carried out a photometric study of the spatial
distribution of the objects in the data release six of the SDSS (SDSS-DR6), and found evidence of an
overdensity in the region of the decrement which is intermediate between randomly selected intercluster regions and galaxy
clusters in the CrB-SC. This overdensity stems chiefly from an excess of faint and red galaxies, which presents a radial
profile extending up to $\sim 15$~arcmin, and could be due to a galaxy cluster with a lack of bright galaxies at the 
CrB-SC redshift ($z\sim 0.07$) or, alternatively, to the brightest part of the luminosity function of a background cluster.
Confirmation and delineation of such structures, and an estimation of the amplitude of the tSZ effect expected from 
them, could be achieved by an spectroscopic study. 

This is addressed in the present paper by means of SDSS-DR7 spectroscopic data.
In section~2 we describe the selection of galaxies in the SDSS-DR7, while in section~3 we present the analyses performed 
on these data intended at analysing whether there is a significant overdensity at the position of CrB-H in the form of either 
a collapsed structure or an elongated filament. In section~4 we make estimates of the tSZ effect and the X-ray flux from 
the gas distribution that the galaxies in that position could be tracing. In section~5 we consider other possible 
CMB secondary anisotropies arisen in more distant and yet undetected massive structures in the light of sight. 
Our main conclusions are drawn in section~6.

\section{SDSS-DR7 data}

With the SDSS-DR7, released in November 2008, the SDSS has obtained spectroscopic data over more than 1,6 million objects,
including 930,000 galaxies, 120,000 quasars and 460,000 stars, with a magnitude limit $r=17.77$ and covering a total sky 
area of 9,380 square degrees
\citep{abazajian_08}. The region of the CrB-SC was fully covered after this release. We built a catalogue of galaxies with
spectroscopic measurements in this
region, encompassing equatorial coordinates $225.0\degr\le{\rm R.A.}\le 237.5\degr$, $25.0\degr\le{\rm Dec.}\le 33.5\degr$,  
from the data releases 4 to 7. To this aim, we downloaded from the SDSS website\footnote{http://www.sdss.org/} the fits
files corresponding to these four data releases. To avoid redundancy, we cross-correlated the previous files and considered
any pair of entries with coordinates separated by less than 0.1'' to be the same object.

The final catalogue in the region of CrB-SC contains 11,842 galaxies with spectroscopy. From all the physical information
available, we kept only that relevant to this study, i.e. the equatorial coordinates R.A. and Dec. (J2000), the photometric
magnitudes in bands $u$, $g$, $r$, $i$ and $z_{f}$, and the spectroscopic redshift with its error.

\section{Galaxy distribution toward CrB-H}

\begin{figure*}
\includegraphics[width=17.0cm]{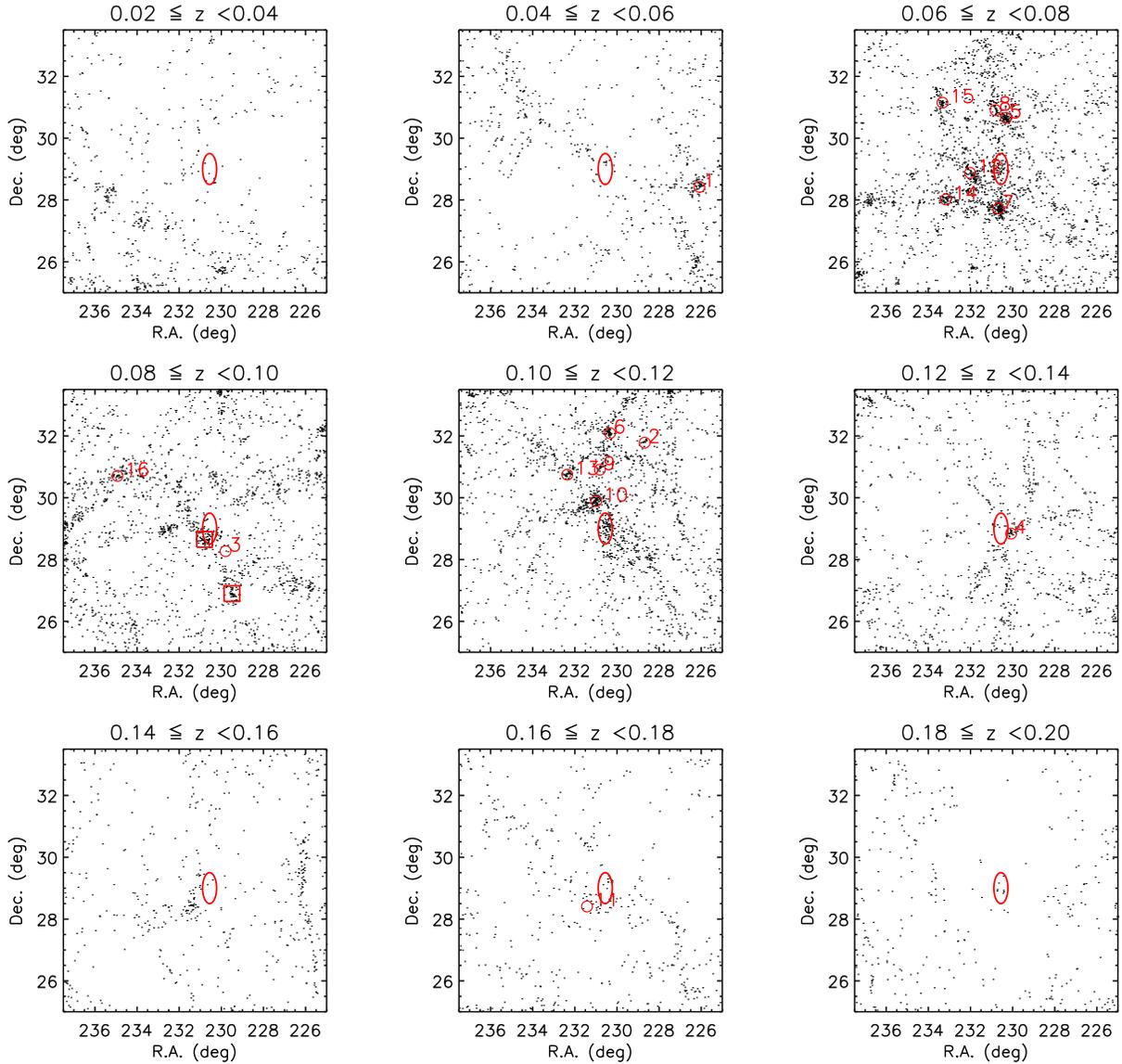}
\caption{Spatial distribution of SDSS-DR7 galaxies in the CrB-SC region, for different $\Delta z=0.02$ redshift slices. 
Circles with labels indicate the positions of Abell clusters in each redshift bin in the region 
(1. Abell 2022; 2. Abell 2049; 3. Abell 2056; 4. Abell 2059; 5. Abell 2061; 6. Abell 2062; 7. Abell 2065; 8. Abell 2067; 
9. Abell 2067B; 10. Abell 2069; 11. Abell 2073; 12. Abell 2079; 13. Abell 2083; 14. Abell 2089; 15. Abell 2092 and 16. 
Abell 2110), the ellipse the location of the CrB-H decrement detected in VSA observations, and the squares two notable 
overdensities outside Abell clusters in the redshift slice $0.08<z<0.10$.
}
\label{fig:spatial_distr}
\end{figure*}

Figure~\ref{fig:spatial_distr} shows the spatial distribution of galaxies in the area of the sky covered by the CrB-SC 
in $\Delta z=0.02$ slices. A remarkable galaxy overdensity is visible at the redshift of the CrB-SC, $z\approx 0.07$. The 
galaxies belonging to the CrB-SC clusters A2061, A2065, A2067, A2079, A2089 and A2092 are clearly seen in this redshift interval. 
Also, a great
concentration of galaxies is notable at $z\approx 0.11$, the redshift of the galaxy cluster A2069. Some authors (e.g. 
Postman et al. 1988) have previously noticed this galaxy overdensity, suggesting the existence of the so-called 
`A2069 supercluster'.

At the position of the decrement there
is also a clear overdensity of galaxies in the bins $z=0.07$ and $z=0.11$, but it does not hold at intermediate redshifts.
Therefore, from these plots it is not evident the existence of a filamentary structure extended toward the line of sight
at the position of the decrement.

\subsection{Galaxy population}

\begin{figure*}
\begin{center}
\includegraphics[width=7.0cm]{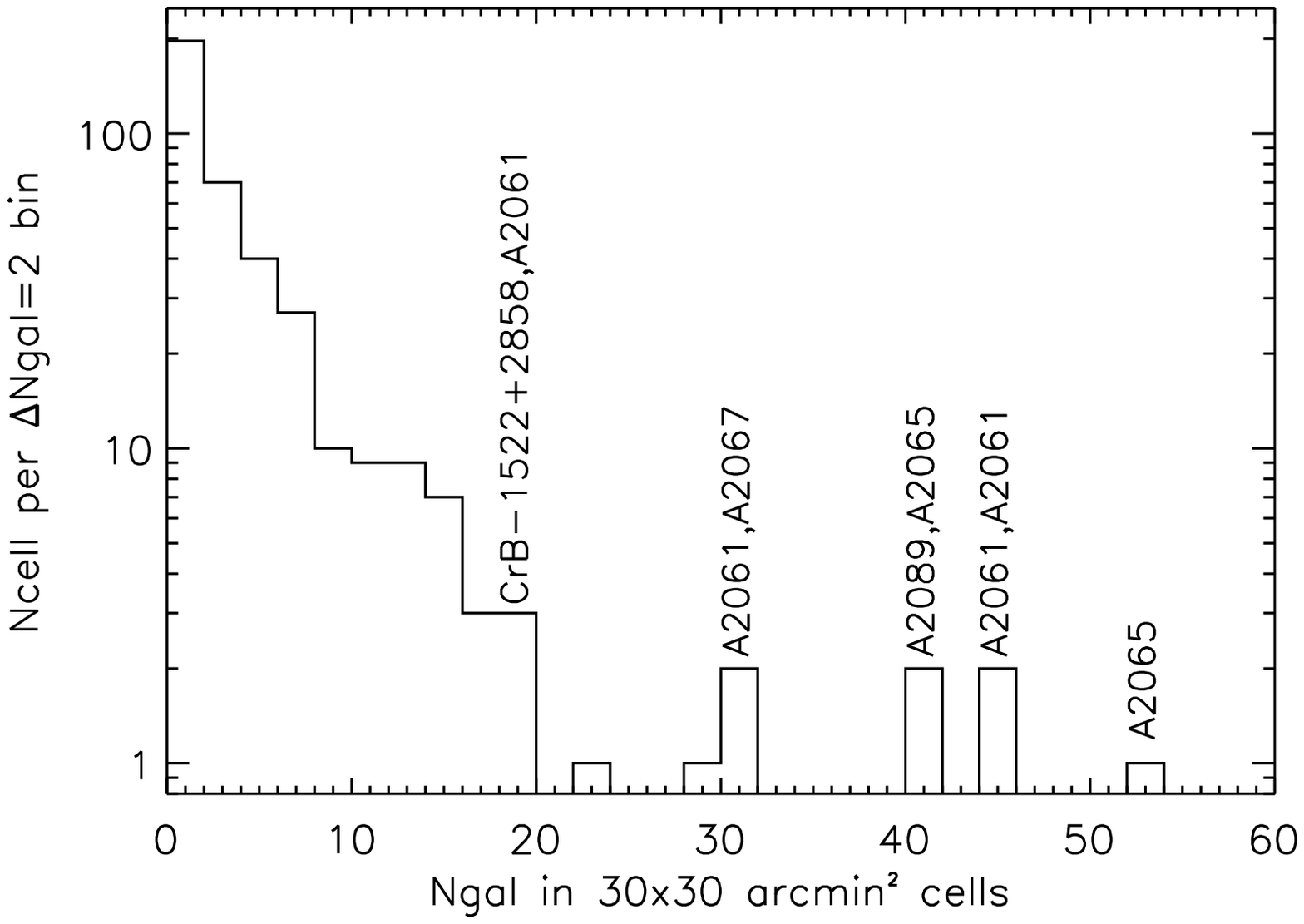}%
\includegraphics[width=7.0cm]{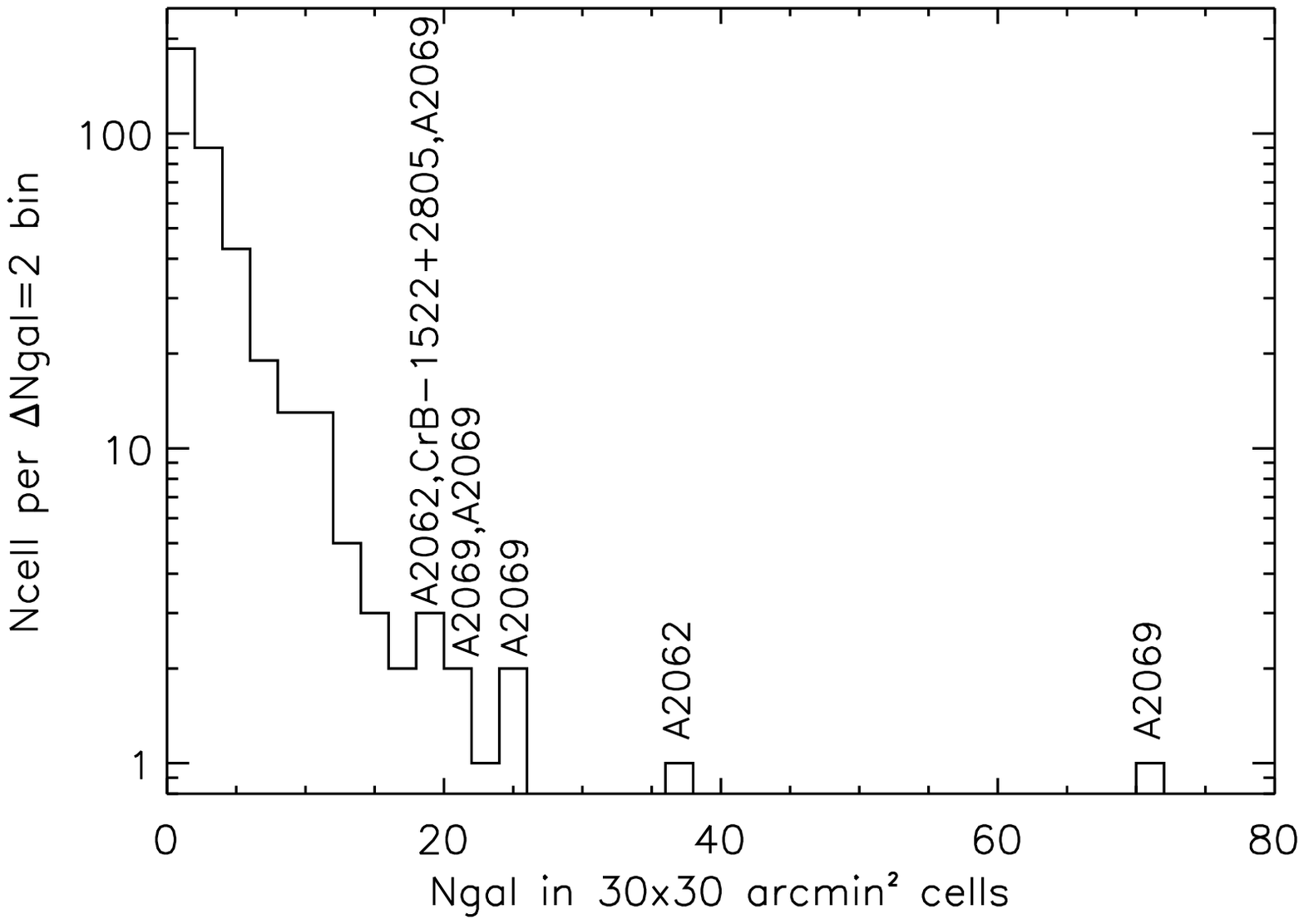}
\end{center}
\caption{Histograms showing the distribution of number of cells of angular size $30\times 30$~arcmin$^2$ and redshift depth
$\Delta z$=0.01 against galaxy population inside the cell, for the redshift slices $0.07<z<0.08$ (left) and $0.11<z<0.12$
(right). The cells at the position of CrB-H have both N$_{\rm gal}=18$ galaxies. Most of denser cells are associated with 
the Abell galaxy clusters indicated by labels on top of each histogram bar.}
\label{fig:histo_z_slides}  
\end{figure*}

In order to assess how significant is the overdensity found at the position of the CrB-H decrement, we have gridded a
volume of angular size $225.3\degr<$R.A.$<237.3\degr$ by $25.3\degr<$Dec.$<33.3\degr$ and depth $0.05<z<0.12$ into 2702 
cells of size
30$\times$30~arcmin$^2~\times~\Delta z=0.01$. In Table~\ref{tab:table_1} we show the forty pixels with the highest number of
SDSS-DR7 galaxies between $z=0.05$ and $z=0.12$, together with the number of galaxies in each of the seven cells
corresponding to each pixel. Numbers labelled with `c' indicate galaxies belonging to any known Abell cluster. In the
penultimate column we show the Abell clusters in the region with coordinates within each particular pixel, whereas in the
last column we indicate other Abell clusters closer than 40' to the pixel coordinates which are having an important
contribution to the projected densities. The pixel corresponding to the CrB-H is the sixteenth pixel with the highest number
of galaxies in the considered redshift range, most of which lie around redshifts $z=0.075$ and $z=0.115$. 
Note that all pixels with a
higher galaxy density are associated with known galaxy clusters. For instance, the pixel with the highest number of galaxies
at $0.05<z<0.12$, which has central coordinates R.A.=$231.06\degr$, Dec.=$30.00\degr$, has 70 galaxies in the cell centred
at $z=0.115$ which belong mostly to the galaxy cluster A2069. The CrB-H pixel is therefore the densest outside Abell 
clusters. Remarkably, it presents notable overdensities at two distant redshifts;
it has 18 galaxies in the cell centred at $z=0.075$ and 18 galaxies in the cell at $z=0.115$. At these
positions \citet{gal_03} and \citet{koester_07} found overdensities of galaxies in the Second Palomar Observatory Sky Survey
and in the SDSS data respectively, and suggested the presence of candidate galaxy clusters. Note that such density of 18
galaxies per cell is rather uncommon outside Abell clusters. 
The same conclusion is drawn from Figure~\ref{fig:histo_z_slides}, where we show the distribution 
of number of galaxies per cell. In the redshift slices $0.07<z<0.08$ and $0.11<z<0.12$ only 2.6\% and 2.3\% of the cells are
found with more than 18 galaxies, respectively. For $0.07<z<0.08$ most of the isolated peaks in the histogram with 
N$_{\rm gal}>20$ are associated with the galaxy clusters A2065, A2061, A2089 and A2067, whereas for $0.11<z<0.12$ these peaks are 
associated with A2069 and A2062. It is then clear that the CrB-H cells are at the high density tail of the distribution of cells 
outside galaxy clusters.

In the table we found only eight cells with more than 18 galaxies outside Abell clusters. 
It is worth noting that two of these cells lie at the intermediate redshift  
$z\approx 0.085$. One of them is centred at R.A.=$229.56\degr$, Dec.$=27.00\degr$ and encloses 34 galaxies in the 
redshift bin $0.08<z<0.09$, and the other is centred at R.A.=$230.56\degr$, Dec.=$28.50\degr$, right south of 
CrB-H, and contains 21 galaxies in the same redshift bin, with other 13 galaxies in the adjacent bin. These overdensities 
are noticeable in Figure~\ref{fig:spatial_distr}, at positions (indicated by the two squares) close to the line connecting 
A2065, the most massive cluster in CrB-SC, and the large background overdensity around A2069. This could be an 
indication of a possible connection between these two structures. However, right at the position of CrB-H there is a lack of 
galaxies at intermediate redshifts.

\begin{table*}
\begin{minipage}{170mm}
\begin{center}
\caption{Forty regions with the highest density of galaxies with redshifts $0.05<z<0.12$ in the CrB-SC area. 
Each region is delineated by a cell with a projected size of $30\times 30$~arcmin$^2$ and a depth of $\Delta z=0.01$. The sky 
coordinates of each cell are quoted in the first two columns. Columns running from the third to the ninth indicate, for each
pixel, the number of galaxies contained in cells of redshifts from $z=0.055$ to $z=0.115$. Numbers labelled with `c' 
indicate galaxies belonging to Abell clusters.  
In the column next to the last we quote the Abell clusters with coordinates lying within each particular pixel. In the last
column we show Abell clusters closer than 40' to the pixel centre coordinates, and their separations inside brackets. The
last two rows show the average number of galaxies per cell in each redshift slice, considering respectively all galaxies and
only galaxies outside Abell clusters.
\label{tab:table_1}}
\scriptsize{
\begin{tabular}{cccccccccccc}
\hline
RA & Dec. &\multicolumn{7}{c}{N$_{\rm gal}$ per $\Delta z=0.01$ interval} &  \\
\noalign{\smallskip}
\cline{3-9}
\noalign{\smallskip}
\noalign{\smallskip}
&& 0.05-0.06 & 0.06-0.07 & 0.07-0.08 & 0.08-0.09 & 0.09-0.10 & 0.10-0.11 & 0.11-0.12 && \\ 
\hline 
231.06 &  30.00  &  0 &   5 &	7 &   3 &   0 &  13 &  70$^{\rm c}$ & A2069 &\\
230.56 &  27.50  &  0 &  22$^{\rm c}$ &  52$^{\rm c}$ &   2 &   0 &   1 &	6& A2065 &\\
230.56 &  28.00  &  0 &  15$^{\rm c}$ &  41$^{\rm c}$ &   3 &   0 &   5 &  10&   &A2065 (18')\\
230.06 &  30.50  &  0 &   5 &  44$^{\rm c}$ &  13 &   0 &   0 &	4 &   &A2061 (16')\\
230.56 &  30.50  &  0 &   0 &  44$^{\rm c}$ &  11 &   2 &   0 &	4$^{\rm c}$ & A2061 & A2067 (28')\\
232.06 &  29.00  &  0 &  46$^{\rm c}$ &	1 &  10 &   1 &   0 &	3 & A2079  &\\
231.56 &  29.50  &  7 &   4 &	3 &  16 &   0 &   9 &  21$^{\rm c}$ &   & A2069 (38')\\
230.56 &  31.00  &  0 &   0 &  30$^{\rm c}$ &   6 &   1 &   2 &  17$^{\rm c}$&   &A2061 (24'), A2067 (14')\\
226.06 &  28.50  & 37$^{\rm c}$ &  13 &	1 &   2 &   0 &   1 &	1 &  A2022 &\\	
230.56 &  32.00  &  0 &   1 &	3 &   0 &   5 &   8 &  37$^{\rm c}$ & A2062 &\\
232.56 &  29.00  &  1 &  13$^{\rm c}$ &	1 &  33 &   1 &   0 &	5&   & A2079 (29')\\
231.56 &  30.00  &  2 &   5 &	2 &  13 &   0 &   3 &  24$^{\rm c}$ &   & A2069 (30')\\
233.06 &  28.00  &  0 &   6 &  41$^{\rm c}$ &   1 &   0 &   0 &	1 & A2089 &\\
233.56 &  31.50  &  1 &  20$^{\rm c}$ &	2 &   6 &   1 &  13 &	5&   & A2092 (24')\\
233.56 &  31.00  &  4 &  33$^{\rm c}$ &	0 &   1 &   6 &   2 &	2 & A2092 &\\
230.56 &  29.00  &  4 &   1 &  18 &   5 &   0 &   1 &  18 & CrB-H &\\
232.06 &  30.50  &  2 &  10 &	6 &  12 &   7 &   6 &	3$^{\rm c}$ &   & A2083 (21')\\
229.56 &  27.00  &  5 &   1 &	3 &  34 &   2 &   0 &	1 &   &\\
231.06 &  31.00  &  0 &   2 &  30$^{\rm c}$ &   1 &   0 &   1 &  11$^{\rm c}$ & A2067 &\\
230.56 &  28.50  &  0 &   1 &  13 &  21 &   2 &   1 &	7 &   &\\
228.56 &  28.00  &  2 &   1 &	6 &   6 &   6 &   0 &  24 &   &\\
232.06 &  29.50  &  4 &  14$^{\rm c}$ &	0 &   9 &   1 &   0 &  15&   & A2079 (37')\\
231.56 &  28.50  &  0 &  33$^{\rm c}$ &	4 &   0 &   3 &   1 &	2&   & A2079 (33')\\
230.56 &  29.50  &  1 &   0 &  15 &   2 &   0 &   3 &  21$^{\rm c}$ &   & A2069 (32')\\
231.56 &  29.00  &  0 &  18$^{\rm c}$ &	7$^{\rm c}$ &  14 &   1 &   0 &	1 &   & A2079 (25')\\	
230.06 &  28.00  &  0 &   4 &  16$^{\rm c}$ &  10$^{\rm c}$ &   0 &   5 &	6&   & A2065 (37'), A2056 (21')\\
234.56 &  31.00  & 11 &   4 &	0 &   4 &  17$^{\rm c}$ &   1 &	1 &   & A2110 (26')\\	
236.56 &  28.00  &  1 &   4 &  22 &  10 &   0 &   0 &	1 &   &\\
237.06 &  28.00  &  0 &   0 &  28 &   3 &   3 &   1 &	2&   &\\
230.56 &  31.50  &  0 &   2 &	6$^{\rm c}$ &   0 &   0 &   9 &  19$^{\rm c}$ &   & A2067 (38'), A2062 (37')\\
232.06 &  28.50  &  0 &  24$^{\rm c}$ &  10$^{\rm c}$ &   1 &   1 &   0 &	0 &   & A2079 (23')\\
231.06 &  28.50  &  1 &   5 &	8 &  15 &   4 &   1 &	2 &   &\\
231.06 &  29.50  &  3 &   2 &	5 &   3 &   0 &   5 &  17&  & A2069 (24') \\
229.56 &  28.50  &  0 &   0 &  15 &   3$^{\rm c}$ &   0 &   2 &  15& A2056 & \\
227.06 &  27.00  &  8 &   1 &  12 &   9 &   0 &   1 &	4&   &\\
229.56 &  33.00  &  1 &   3 &	0 &   2 &   5 &   0 &  23 &  & \\
232.06 &  31.00  &  2 &   1 &	4 &   0 &   0 &  23 &	4 &  & A2083 (22')\\
230.06 &  31.00  &  0 &   3 &  18$^{\rm c}$ &   0 &   2 &   1 &  10$^{\rm c}$ &   & A2061 (24'), A2067 (39')\\
230.56 &  30.00  &  0 &   2 &  11 &   0 &   0 &   2 &  19$^{\rm c}$ &   & A2069 (23')\\
231.06 &  29.00  &  1 &   1 &  15 &  13 &   0 &   0 &	4 &   & \\
\hline
\multicolumn{2}{c}{Average} & 1.38 & 2.50 & 3.74 & 3.05 & 1.58 & 1.56 & 3.31 \\
\multicolumn{2}{c}{Average no clusters} &  0.99 & 1.25 & 1.71 & 1.77 & 1.07 & 0.85 & 1.36 \\
\hline
\end{tabular}
}
\end{center}
\end{minipage}
\end{table*}

\subsection{Redshift distribution}

\begin{figure*}
\begin{center}
\includegraphics[width=5.0cm]{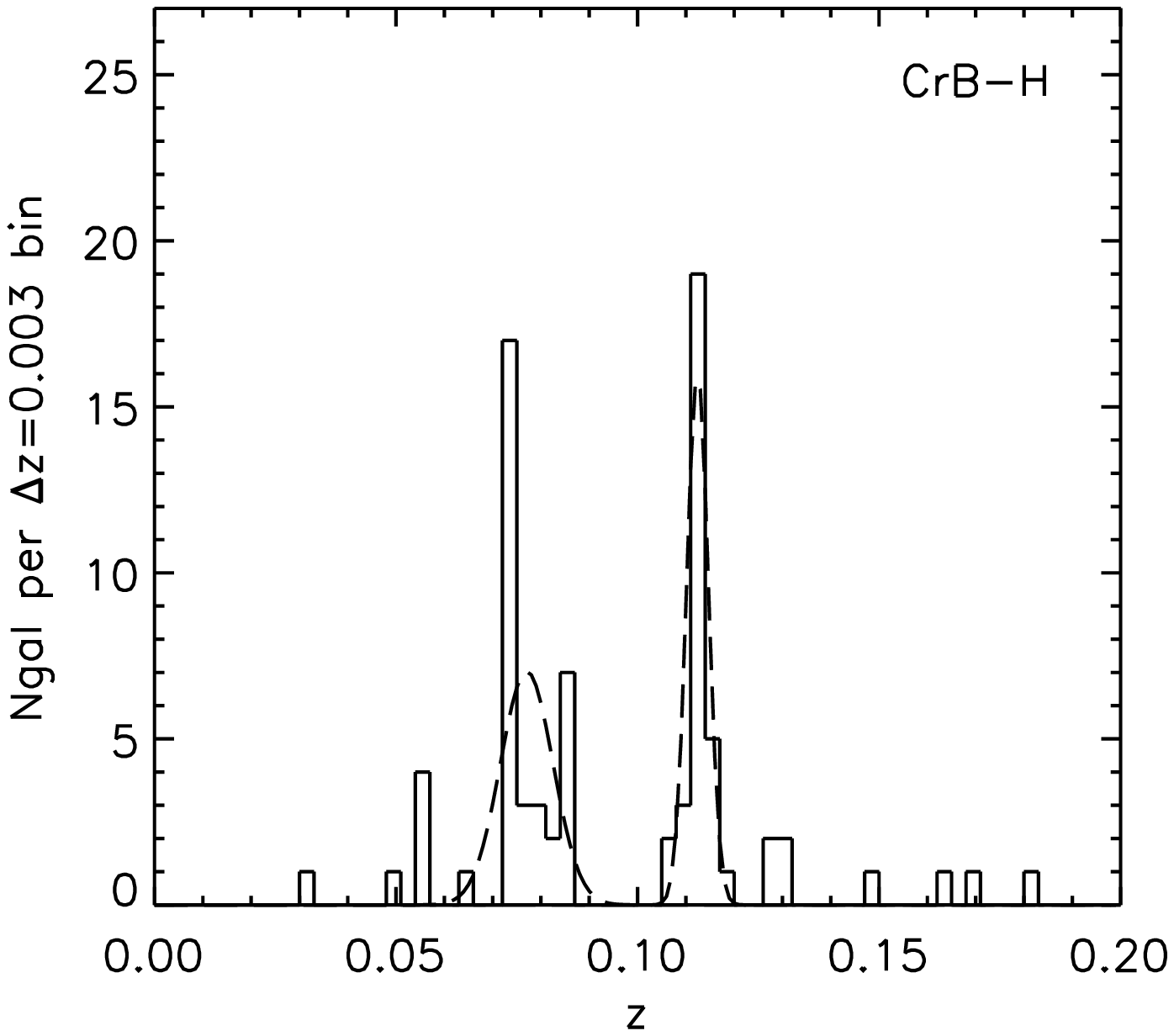}%
\includegraphics[width=5.0cm]{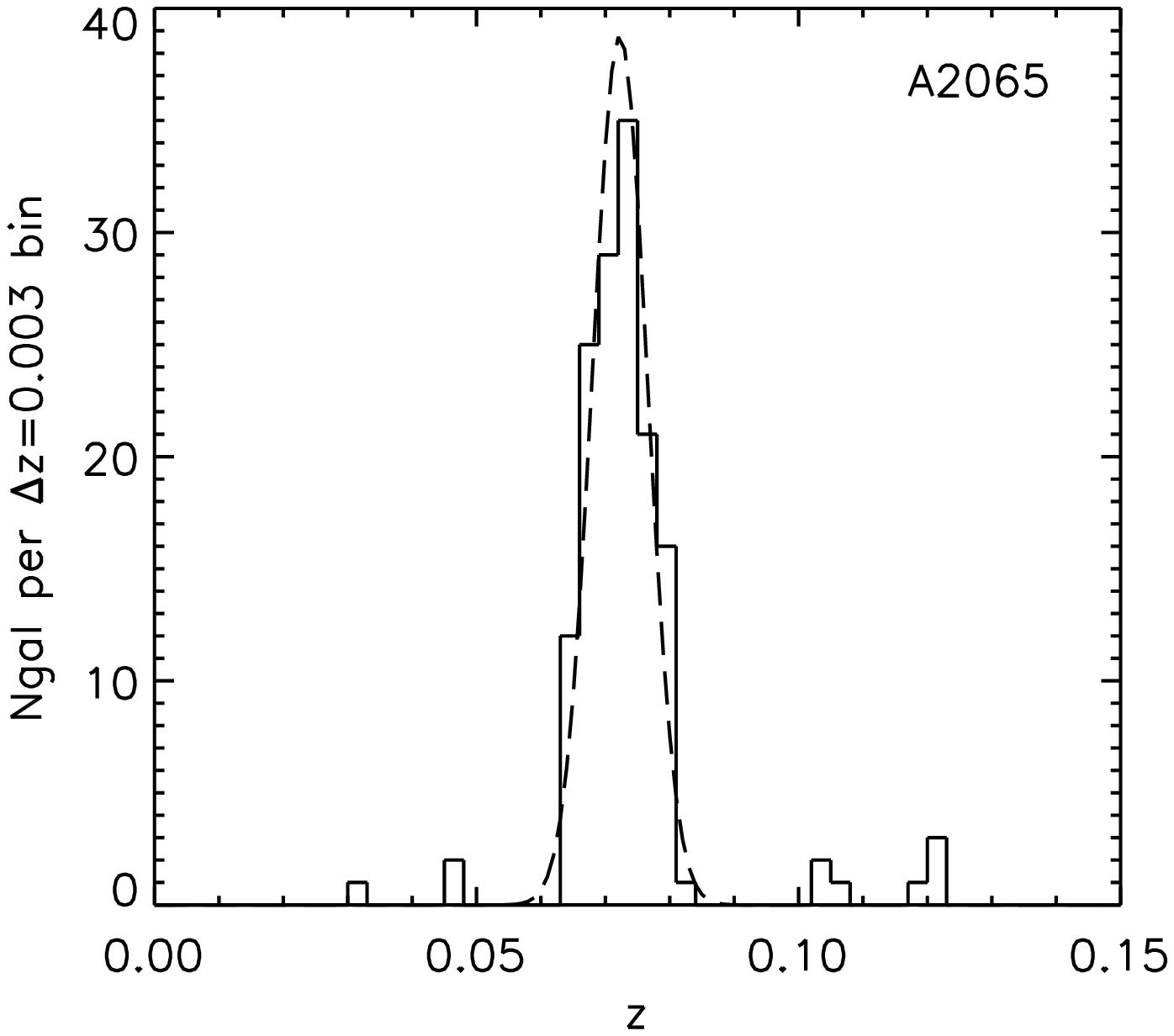}%
\includegraphics[width=5.0cm]{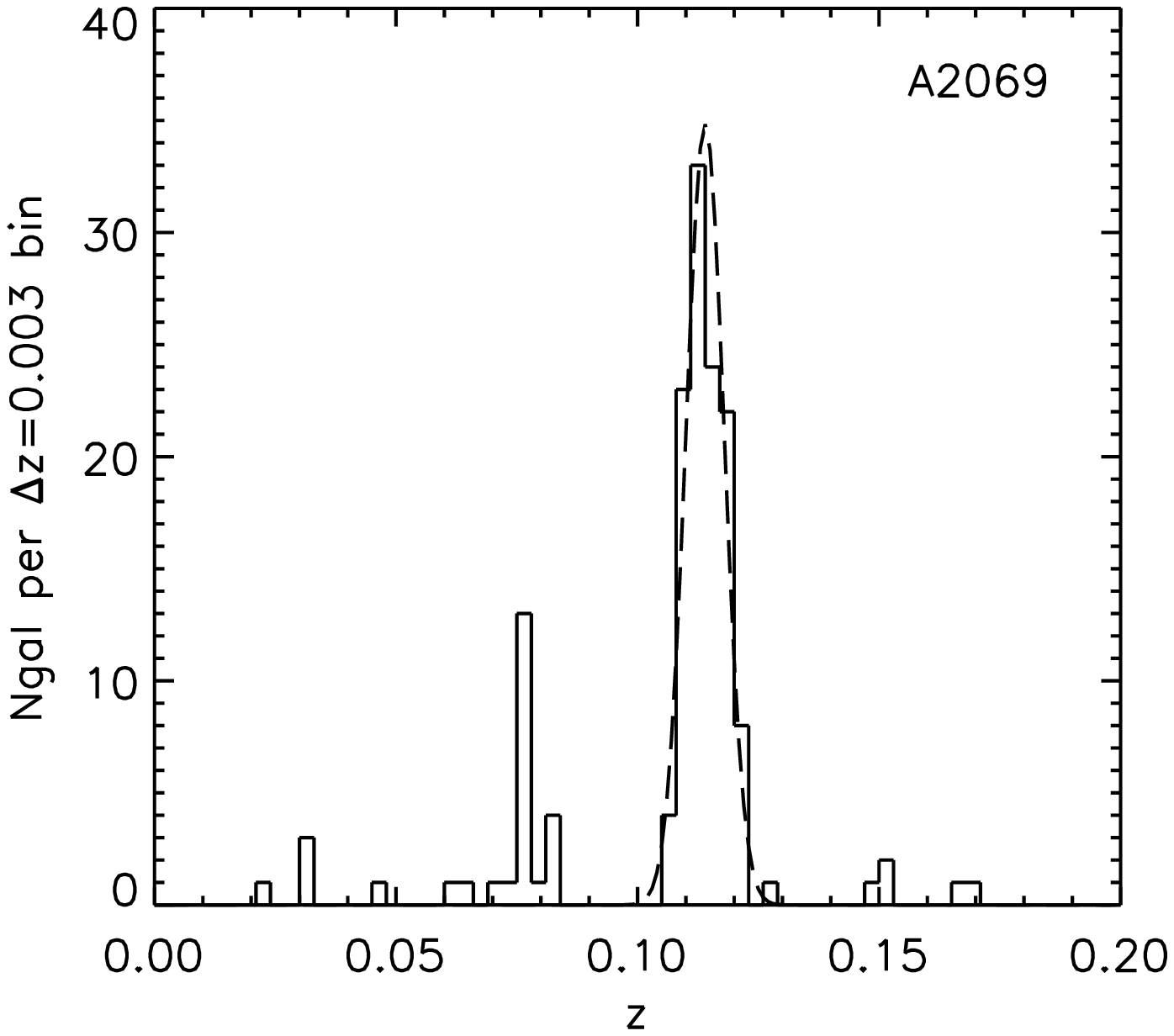}
\end{center}
\caption{Redshift distribution of galaxies within 20' of the CrB-H decrement (left), in comparison with those of the galaxy
clusters A2065 (centre) and A2069 (right). The dashed lines show Gaussian fits to the galaxy distribution in each case.
}
\label{fig:z_histograms}  
\end{figure*}

The redshift distribution of galaxies within 20' of the CrB-H decrement is shown in Figure~\ref{fig:z_histograms}. There are two clear 
peaks in this distribution at $z\approx 0.07$ and $z\approx 0.11$, which are indicative of the
two overdensities found in the redshift bins $0.07<z<0.08$ and $0.11<z<0.12$ in Table~\ref{tab:table_1}.
This histogram can be compared, in the same figure, with the 
redshift distribution of galaxies toward the direction of the two richest clusters in the region: A2065, which belongs to the 
CrB-SC, and the more distant cluster A2069. 
The mean redshift and standard deviations of all galaxies associated with the two peaks in the CrB-H decrement 
histogram, and with the single peaks in the A2065 and A2069 histograms, are shown in Table~\ref{tab:table_2}. 
We have identified respectively 33 and 30 galaxies
within a radius of 20' around the CrB-H decrement, and with redshifts $0.06<z<0.10$ and $0.10<z<0.12$, which we associate with these
galaxy overdensities. The centroid of these galaxies have coordinates 
RA=15$^{\rm h}$22$^{\rm m}$21$^{\rm s}$, Dec.=28$^\circ$58'32'' and RA=15$^{\rm h}$22$^{\rm m}$00$^{\rm s}$, Dec.=28$^\circ$05'10''
respectively, and therefore we name these overdensities as CrB-1522+2858 and CrB-1522+2805.

From the redshift dispersions (which have been estimated directly from the individual galaxy redshifts and not from the histogram), 
and from the mean error of the SDSS redshift estimates ($z_{\rm err}\approx 0.0002$), we have computed the radial velocity dispersions and their 
errors following the formalism described in \citet{danese_80}.  We approximate the virial mass by $M_{200}$, i.e. the mass within 
$r_{200}$, the 
radius inside which the density is 200 times the critical density. By assuming virial equilibrium, and writing the 3D velocity dispersion as a function 
of the radial velocity dispersion, $(\sigma_v^2)_{\rm 3D}=3\sigma_v^2$ \citep{abell_77}, we estimate $M_{200}$ and $r_{200}$ (see explicit 
equations for example in Finn et al. 2005 or in D{\'{\i}}az-S{\'a}nchez et al. 2007). We also compute the electron temperature of these structures, 
under the assumption that the intracluster gas shares the same dynamics as member galaxies (equation~3 of Rosati et al. 2002).
In table~\ref{tab:table_2} we show the radial velocity dispersions, $M_{200}$ masses and electron temperature for 
CrB-1522+2858, CrB-1522+2805, and for the galaxy clusters A2065 and A2069, which were computed using WMAP5 cosmology:
$\Omega_m=0.26$, $\Omega_\Lambda=0.74$, $h=0.72$ \citep{dunkley_09}. The values for A2065 and A2069 are consistent with 
other estimates. \citet{struble_99} give
$\sigma_v=1203$~km~s$^{-1}$ for A2065. 
\citet{reiprich_02} and \citet{brownstein_06} estimated masses for A2065 of 
$M_{200}=23.37^{+29.87}_{-9.42}\times 10^{14}~h_{50}^{-1}$~M$_\odot$ and 
$M_{250}=8.01^{+7.20}_{-2.99}\times 10^{14}$~M$_\odot$, respectively. 
\citet{small_98} found
$\sigma_v=1203^{+371}_{-289}$~km~s$^{-1}$ and $M_{\rm vir}=18.5^{+13.2}_{-8.2}\times 10^{14}~h^{-1}$~M$_\odot$ for A2065.
For the A2069 supercluster they give $\sigma_v=1684^{+145}_{-151}$~km~s$^{-1}$ and $M=6\times 10^{16}~h^{-1}$~M$_\odot$, but
they consider a much wider volume than our, that encloses a total of 352 galaxies. 

The values obtained for the velocity dispersion, mass and temperature of CrB-1522+2858 are so high that the hypothesis of a
bound structure seems unrealistic. For the higher-redshift structure, CrB-1522+2805, the values are however within the range of
typical galaxy clusters. In Figure~\ref{fig:z_histograms} it is also evident that the CrB-1522+2805 peak is narrower and
steeper, and more similar to galaxy cluster than the CrB-1522+2858 one. Therefore, according to this analysis, the 
higher-redshift structure could be virialized to some extent.

\begin{table*}
\begin{minipage}{170mm}
\begin{center}
\caption{Means and standard deviations of the redshifts, and inferred velocity dispersions, for galaxies in the listed 
redshift ranges and within 20' of the CrB lower-redshift and higher-redshift structures, and within the same radius of the 
galaxy clusters A2065 and A2069. We also show the derived masses and electron temperatures.\label{tab:table_2}}
\begin{tabular}{cccccccc}
\hline
 &  $z_{\rm min}-z_{\rm max}$ & N$_{\rm gal}$ & $\bar{z}$ & $\sigma_z$ & $\sigma_v$ (km~s$^{-1}$) & $M_{200}$ 
 (10$^{15}$~M$_{\odot}$) & $k_{\rm B} T_{\rm e}$ (keV)\\
\hline
CrB-1522+2858  &  0.06-0.10   & 33 & 0.0770 & 0.0054 & 1514$^{+231}_{-158}$$^*$ & 5.65$^{+2.59}_{-1.77}$ $^*$ & 14.2$^{+4.3}_{-3.0}$ $^*$ \\
\noalign{\smallskip}
CrB-1522+2805 &  0.10-0.12   & 30 & 0.1124 & 0.0022 &  599$^{+97}_{-65}$ & 0.35$^{+0.17}_{-0.11}$ &  2.2$^{+0.7}_{-0.5}$ \\  
\noalign{\smallskip}
A2065          &  0.052-0.092 &139 & 0.0722 & 0.0043 & 1199$^{+79}_{-66}$ & 2.81$^{+0.56}_{-0.47}$ &  8.9$^{+1.2}_{-1.0}$ \\
\noalign{\smallskip}
A2069          &  0.095-0.125 &114 & 0.1140 & 0.0039 & 1061$^{+78}_{-64}$ & 1.91$^{+0.42}_{-0.35}$ &  7.0$^{+1.0}_{-0.8}$ \\
\hline
\end{tabular}
\end{center}
$^*$ Note that these values are too high and meaningless, as they have been computed under the assumption of virial equilibrium. The
galaxies in this case may not be gravitationally linked, and therefore the velocity dispersion calculated in these cases may be dominated by 
the Hubble flow rather than by peculiar velocities.
\end{minipage}
\end{table*} 

\subsection{Radial distribution}

We have analyzed the radial distribution of the galaxies belonging to the structures CrB-1522+2858 and CrB-1522+2805. We counted
the number of objects lying in concentric rings of increasing radii around the centroid of each structure. The angular profile of 
CrB-1522+2805 is steeper than the CrB-1522+2858 one, but both are clearly flatter than those of A2065 and A2069 clusters, which 
present a much higher number of galaxies in the centre.

\begin{figure*}
\begin{center}
\includegraphics[width=6.5cm]{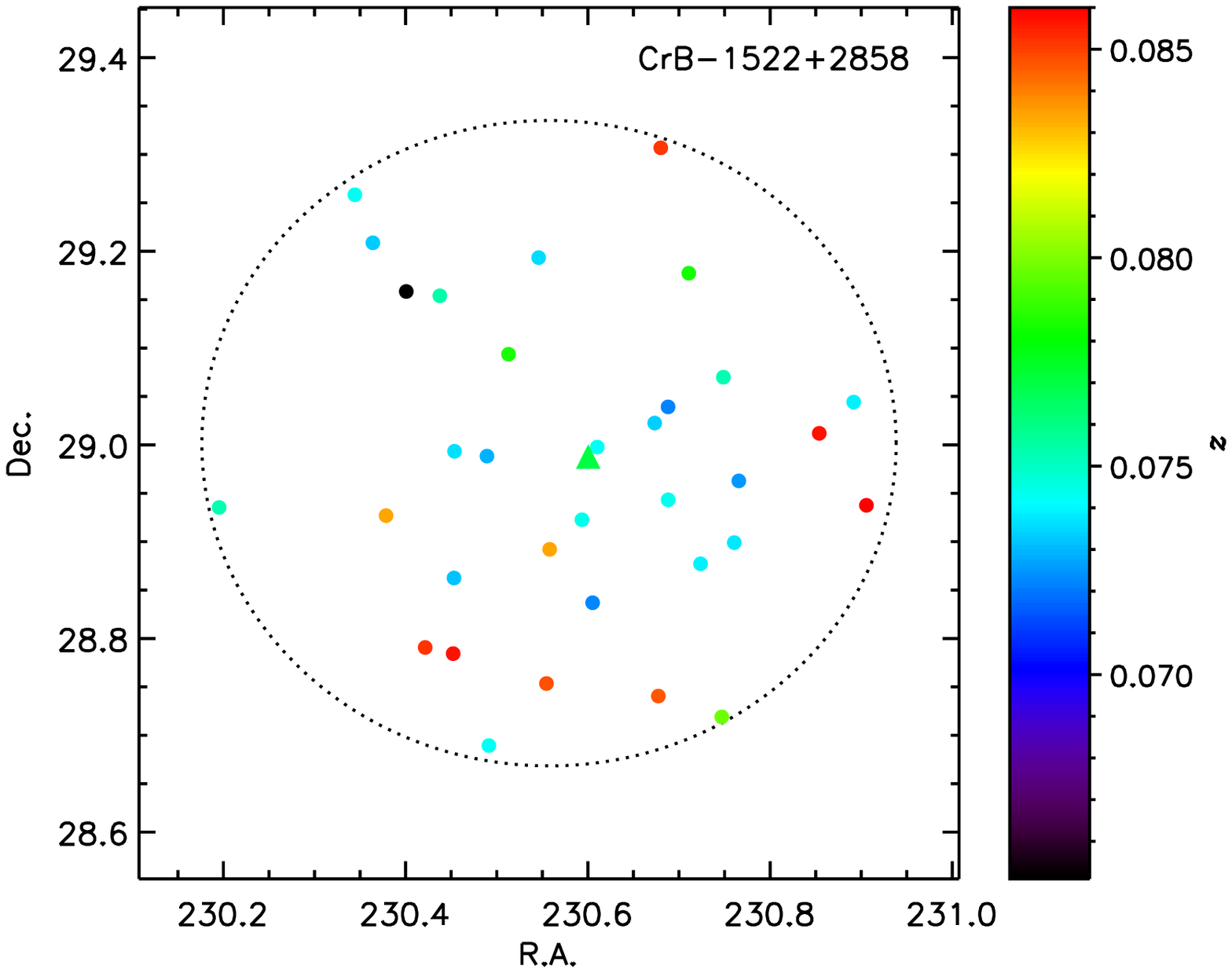}%
\includegraphics[width=6.5cm]{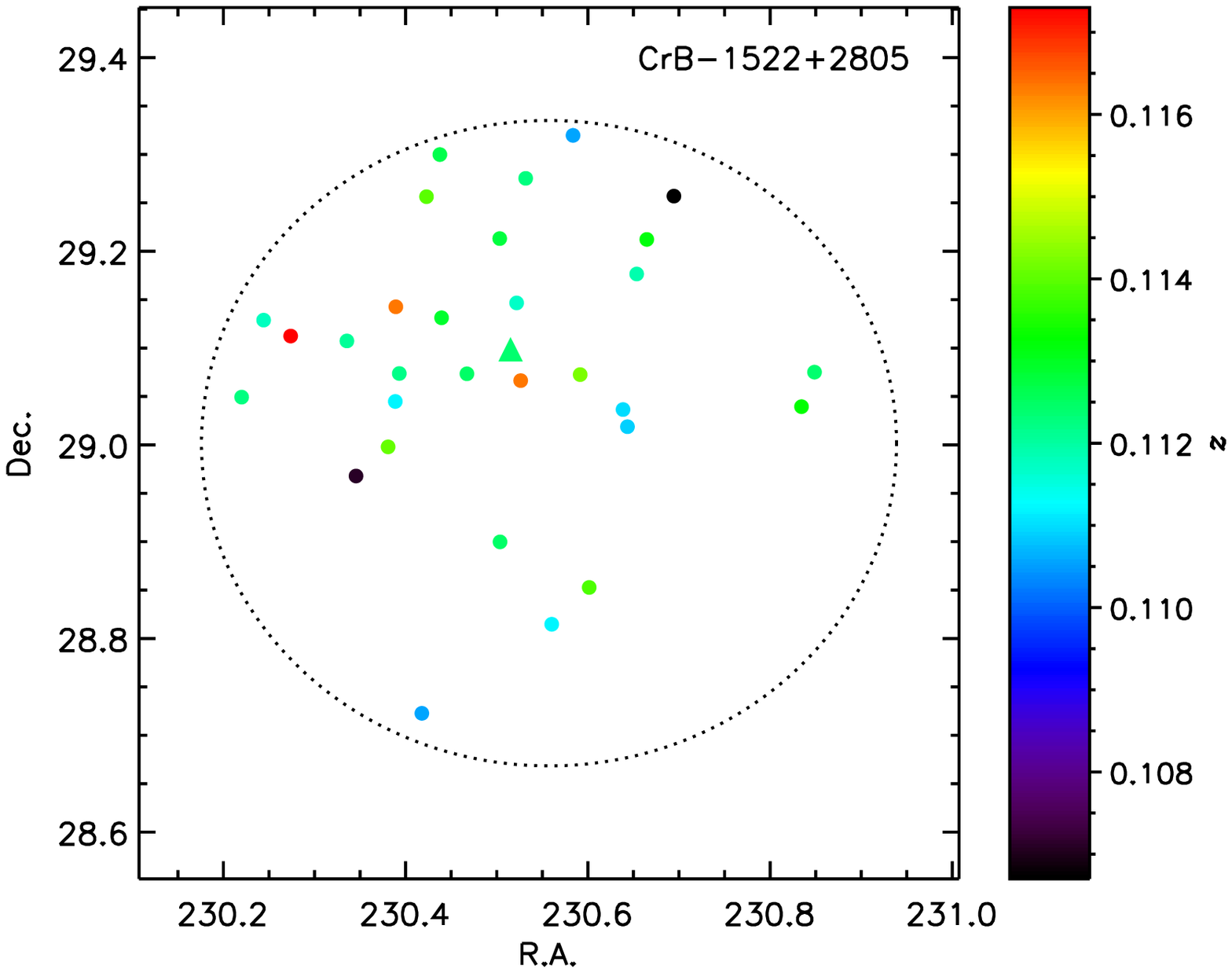}
\includegraphics[width=6.5cm]{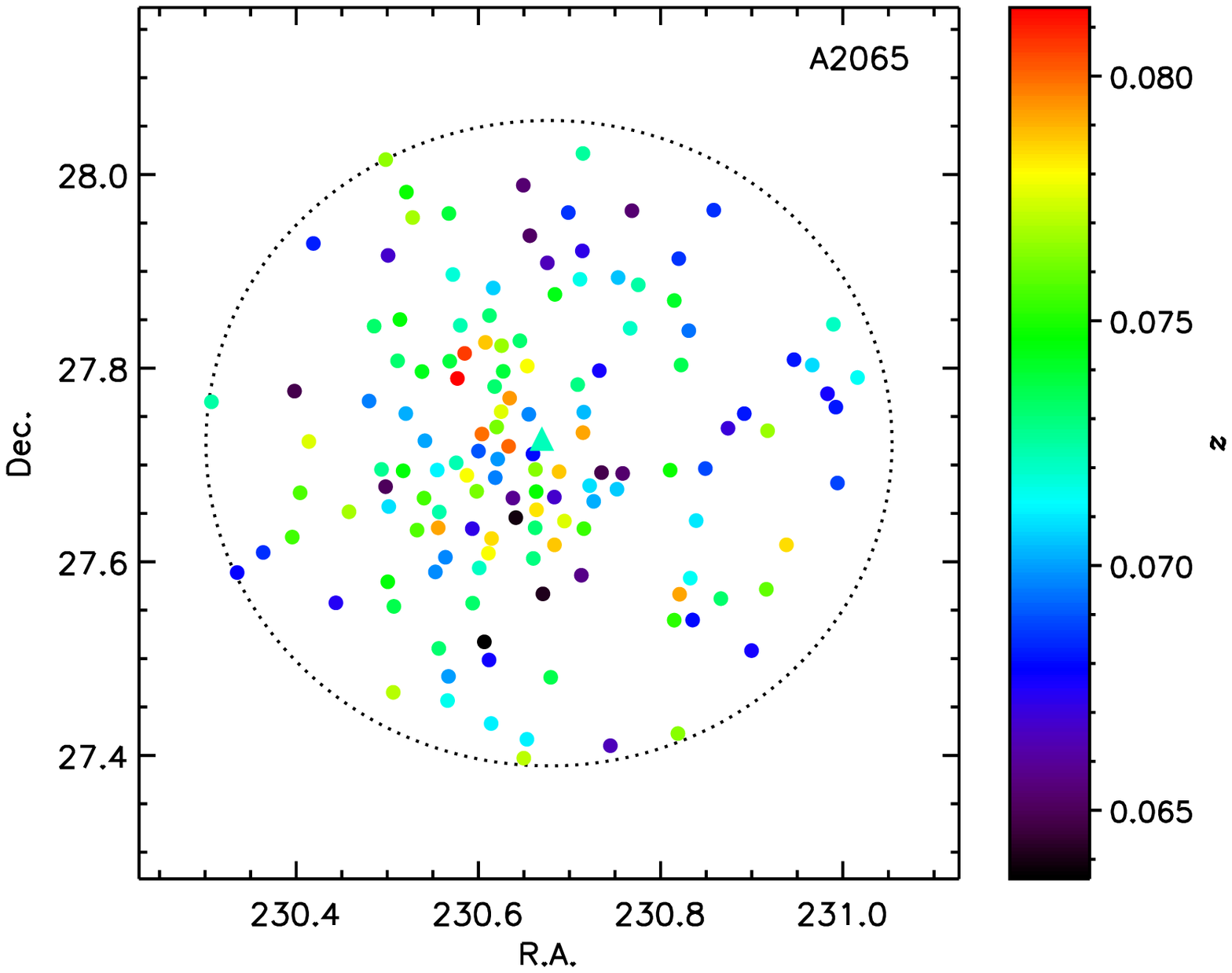}%
\includegraphics[width=6.5cm]{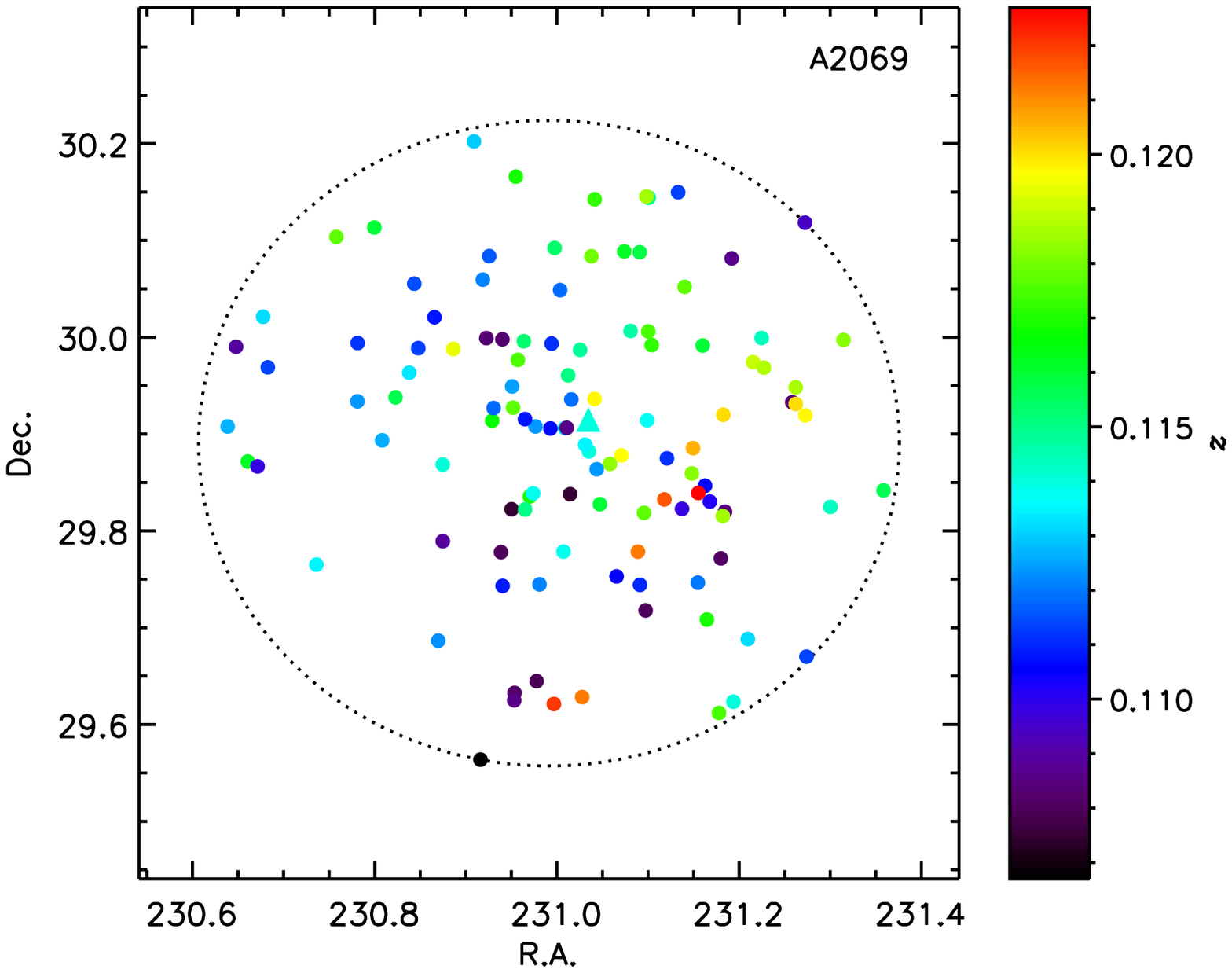}
\end{center}
\caption{Top: spatial distribution of galaxies corresponding to the lower-$z$ (left) and higher-$z$ (right) peaks in the redshift
histogram of galaxies shown in Figure~\ref{fig:z_histograms}. We have considered galaxies within 20' of the CrB-H coordinates, and 
with redshifts $0.06<z<0.10$ and $0.10<z<0.12$, respectively. Bottom: spatial distribution of galaxies within 20' of A2065 (left) 
and A2069 (right), with redshifts
$0.052<z<0.092$ and $0.095<z<0.125$, respectively. Each dot represents a galaxy, and its colour indicate its redshift according to 
the colour scale on the right. The triangle represent the mean coordinates and redshift of all galaxies inside the dotted circle of 
radius 20'.}
\label{fig:spatial_dist}  
\end{figure*}

In Figure~\ref{fig:spatial_dist} we show the spatial distribution of the galaxies around both CrB structures and, for comparison,
around A2065 and A2069. It is notable that the density of galaxies in the clusters peaks in the inner region, whereas in the CrB
structures it is more uniform across the region. Similarly, in the galaxy clusters the redshift dispersion is higher in the centre
than in the external parts, whereas in CrB-1522+2858 and CrB-1522+2805 the galaxy redshifts have similar scatter in the full
region.

In order to quantify the degree of homogeneity of those samples, we
have implemented the following statistical test, commonly known as
``$r^2$-test".  The idea is to use an statistic to test the
null-hypothesis, which in this case would be that the two-dimensional
distribution of galaxies in those regions is drawn from a uniform
distribution. To this end, we calculate the distribution of $r^2$
values, where $r$ is defined as the angular distance of each galaxy to
the centroid of the galaxies in a region of $R=20'$. In the case of a
uniform distribution, the mean value $<r^2>$ should be $R^2/2$, and
the variance $R^4/12/N_{\rm gal}$, being $N_{\rm gal}$ the number of
galaxies in the region.
Table~\ref{tab:stat} shows the results of this analysis when applied
to all cases presented in Figure~\ref{fig:spatial_dist}.
When considering the two Abell clusters (A2065 and A2069), we find
that in both cases the mean value of this statistic deviates
significantly from the expected value for a uniform distribution. For
A2065, we find a $-7\sigma$ deviation, while for A2069, we have
$-6\sigma$.

When applied to the CrB regions, we obtain a $-1.6\sigma$ deviation
for CrB-1522+2858, and $-3.1\sigma$ for CrB-1522+2805. If we do not separate in
redshift, and we study the two-dimension spatial distribution of all galaxies
($0.02<z<0.18$) in the region, we obtain a deviation of $-2\sigma$
(case labelled as CrB-H in the table).
In all these latter three cases, we obtain that the galaxy distribution
is more concentrated than the uniform distribution.  For the CrB-1522+2858,
the distribution is still compatible with a uniform distribution,
which supports the idea that this is a non-virialized object.

However, for the CrB-1522+2805 case, we find some evidence that this galaxy
distribution might trace a virialized structure.
It is important to note that in this case, the significance of the
deviation is smaller than the one for the two clusters, but this is
in part due to the fact that $N_{\rm gal}$ is much smaller here.
The comparison of the actual deviation from the mean value shows
that this CrB-1522+2805 region deviates in the same way and by a similar amount 
as galaxy clusters from the uniform distribution.

\begin{table}
\centering
\caption{Results of the statistical test to probe the spatial distribution of the regions shown
in Figure~\ref{fig:spatial_dist}. First column shows the $<r^2>$-statistic, defined as the
average value of the square of the angular distance of each galaxy to
the centroid of the distribution.  For a uniform distribution, this
value should be $R^2/2$, being $R=20'$ in this case.  Second column
shows the deviation of this value with respect to $1/2$, and the last
column shows the significance of that deviation (see text for
details). }
\begin{tabular}{lccc}
\hline
Region & $\frac{<r^2>}{R^2}$ & $\frac{<r^2>}{R^2}-\frac{1}{2}$ & Significance\\
\hline
A2065 &      0.329 &     -0.171 &      -6.99 \\
A2069 &      0.339 &     -0.161 &      -5.97 \\
CrB-1522+2858 &      0.418 &    -0.0820 &   -1.63 \\
CrB-1522+2805 &      0.336 &     -0.164 &   -3.12 \\
CrB-H &      0.431 &    -0.0688 &      -2.08 \\
\hline
\end{tabular}
\label{tab:stat}
\end{table}

\section{SZ effect and X-ray emission toward CrB-H}
The analyses presented in the previous section revealed the presence of CrB-1522+2805, a possible group 
or low-mass galaxy cluster located at $z\approx 0.11$. On the other hand, there is not any excess in this region of X-ray flux in 
the ROSAT-R6 (0.73-1.56~keV, Snowden et al. 1997) map. In this section we study whether the gas distribution associated with 
CrB-1522+2805 could give rise to a significant SZ signal without  detectable X-ray emission.

\subsection{SZ effect}
\subsubsection{tSZ effect}
The tSZ temperature decrement produced by a galaxy cluster can be expressed as a function of its gas mass, as both are linearly
dependent on the electron density, whose radial dependence can be modelled 
by a $\beta$-profile \citep{cavaliere_76}. The resulting equation is:
\begin{equation}
\Delta T_0 = \frac{3}{2\pi}\frac{T_{\rm CMB} f(x) \sigma_{\rm T} k_{\rm B}T_e}{\mu_e m_e m_p c^2} 
\frac{I(\beta)}{I'(\beta,r/r_c)}\frac{f_{\rm gas}M_{200}}{r_c^2}~~,
\label{eq:dt0}
\end{equation}
where $\mu_e=1.146$ (calculated from the solar abundances of Anders \& Grevesse, 1989) is the mean baryonic mass fraction per 
electron, $I$ and $I'$ are functions of $\beta$ and $r/r_c$ and $f(x)$ is a function of the dimensionless frequency 
$x=\frac{h\nu}{k_{\rm B}T_e}$.

We estimate the expected minimum tSZ decrement of CrB-1522+2805 from the mass and electron temperature of 
Table~\ref{tab:table_2}.  We fix the $\beta$ 
parameter at $\beta=2/3$, and obtain the core radius and the gas fraction from the scaling relations of  \citet{chen_07} 
(see their Table 6) and  \citet{vikhlinin_09} (see their equation 9), respectively. These scaling relations are given as a function of 
$M_{500}$. After rescaling this quantity from $M_{200}$ by assuming hydrostatic equilibrium, we get $r_c=67$~kpc 
and $f_{\rm gas}=0.095$. Introducing these values into equation~\ref{eq:dt0} we obtain a minimum temperature decrement 
at the VSA frequency of $\Delta T_0= -158~\mu$K.

Establishing a reliable comparison between this value and the minimum temperature of $-229~\mu$K found in the VSA map at 
the position of the CrB-H decrement \citep{genova_08}, requires however taking into account the dilution introduced by the convolution 
with the synthesized beam of the VSA observation. This is attained by simulating the tSZ temperature profile of the cluster with the 
model parameters given in the previous paragraph,  and convolving this map with the synthesized beam of the VSA observation. We 
obtain a minimum tSZ temperature decrement in the simulated map of $-26~\mu$K\footnote{Hereafter we will present always temperature 
decrements resulting from the convolution with the VSA beam.}. 

The predicted tSZ effect from CrB-1522+2805 can be compared with that of a typical known galaxy cluster in the region, such as
A2065. \citet{brownstein_06} give for this cluster $\beta=1.162$, $r_c=485.9$~kpc, $r_{250}=1.302$~Mpc, 
$k_{\rm B}T_{\rm e}=5.50$~keV, $M_{\rm gas}=0.49\times 10^{14}$~M$_\odot$ and $M_{250}=8.01\times 10^{14}$~M$_\odot$. 
With these numbers we get a minimum decrement in the VSA map of $\Delta T_0^{\rm A2065}=-107~\mu$K. 
In the initial survey of CrB-SC, with a previous configuration of the VSA having a coarser angular 
resolution, we found a temperature decrement of $-103\pm 56~\mu$K \citep{genova_05} at the position of A2065, which is 
in good agreement with the previous estimate, despite the different angular resolutions.

\subsubsection{kSZ effect}
Coherent movement of the scattering electrons inside the supercluster can build up kSZ effects.
At the VSA frequency, the ratio of the kSZ to the tSZ effect is given by
\begin{equation}
\frac{\Delta T_{0}^{\rm kSZ}}{\Delta T_{0}^{\rm tSZ}} \approx 0.09~\left(\frac{\rm 10~keV}{k_{\rm B} T_{\rm e}}\right) 
\left(\frac{v_{\rm pec}}{1000~\rm km~s^{-1}}  \right)~~.
\end{equation}
For $v_{\rm pec}\sim 1000~{\rm km~s^{-1}}$ and at the electron temperatures typical of galaxy clusters, 
$k_{\rm B}T_{\rm e}\sim 5-10$~KeV, the kSZ effect is an order of magnitude below the tSZ effect. A structure like CrB-1522+2805, 
with a receding peculiar velocity $v_{\rm pec}$, would generate a kSZ effect in the VSA map of 
$\approx -11~\mu$K$\left(v_{\rm pec}/1000~{\rm km~s^{-1}}\right)$. Therefore, if the gas in CrB-1522+2805 have similar properties 
to galaxy clusters or galaxy groups, the contribution from the kSZ effect is unimportant. However, had this structure more special 
characteristics, with a lower temperature typical of WHIM (0.1-1~keV), and a high coherent velocity, it could produce a kSZ signal 
comparable to the tSZ.

\subsection{X-ray flux}

If the electron density is expressed in terms of the gas mass fraction, the X-ray surface brightness from the centre of a galaxy
cluster follows the scaling relation:
\begin{equation}
S_{X0}\propto \frac{T_{\rm e}^{1/2} (f_{\rm gas}M_{200})^2}{(1+z)^4~r_c^5}\frac{F(\beta)}{F'(\beta,r_{200}/r_c)}~~,
\label{eq:Sx0_2}
\end{equation}
where $F$ is a function of $\beta$ and $F'$ of $\beta$ and $r_{200}/r_c$. Using this equation and rescaling to the flux 
in the ROSAT-R6 map at the position of A2065 we can derive the expected signal from CrB-1522+2805 in the same map.
The peak flux toward A2065 is $6.4\times 10^{-4}$~count~s$^{-1}$~arcmin$^{-2}$, being the background
level in this region $\approx 10^{-4}$~count~s$^{-1}$~arcmin$^{-2}$. For the sake of consistency, we will take, for A2065, 
$\beta=1.162$, $r_c=485.9$~kpc, $k_{\rm B}T_{\rm e}=5.50$~keV and $f_{\rm gas}=0.06$ from \citet{brownstein_06}.

Using these parameters for A2065 and those assumed for CrB-1522+2805 in the previous subsection, and considering their
respective redshifts which are 0.0726 \citep{struble_99} and 0.112, equation~\ref{eq:Sx0_2} gives an X-ray flux for 
CrB-1522+2805 of $31\times 10^{-4}$~count~s$^{-1}$~arcmin$^{-2}$.  This high value, which is significantly above the 
background level in the region, arises from the short core radius assumed, $r_c=67$~kpc. However, the $r_c-M_{500}$ scaling 
relation of  Chen et al. (2007; see their Figure~4) allows significantly higher values of $r_c$ for the mass of CrB-1522+2805. As 
we will see in the next section, higher values of $r_c$ produce lower X-ray fluxes.

\subsection{High SZ with low X-ray emission}
\begin{figure}
\begin{center}
\includegraphics[width=\columnwidth]{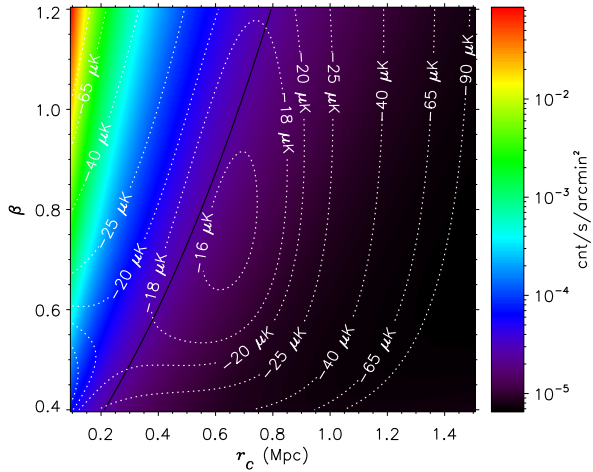}
\end{center}
\caption{Predicted X-ray flux in the ROSAT-R6 data and tSZ effect in the VSA data generated by the CrB-1522+2805 structure 
for different combinations in the parameter space $\beta-r_c$. The logarithmic colour scale show X-ray flux in units of 
count~s$^{-1}$~arcmin$^{-2}$. The solid line marks the ROSAT-R6 background level of 
10~$^{-4}$count~s$^{-1}$~arcmin$^{-2}$. The hypothetical structure should emit below this level, as no detectable X-ray
emission is seen at this position on the map. The white dashed contours indicate the predicted tSZ effect this structure would
imprint in the VSA data.
}
\label{fig:sz_xray_comp}  
\end{figure}

We now explore different combinations of $\beta$ and $r_c$ which could produce a significant tSZ flux with little 
X-ray emission. We initially fix the rest of parameters ($k_{\rm B}T_{\rm e}$, $M_{200}$, 
$f_{\rm gas}$ and $z$) at the same values assumed in section~4.1. The predicted X-ray flux in ROSAT-R6 and tSZ effect in the VSA 
map in the $\beta-r_c$ parameter space are depicted in Figure~\ref{fig:sz_xray_comp}. For core radii 
$r_c\sim 200-400$~kpc, which are within the range allowed by the $r_c-M_{500}$ scaling relation,
we see that the largest tSZ effect we can have without detectable X-ray emission is $\approx -18~\mu$K. We have 
carried out the same analysis when $f_{\rm gas}$ is fixed at different values within the interval 0.08-0.12 (these values 
are allowed by the $f_{\rm gas}-M_{500}$ scaling relation of Vikhlinin et al. 2009; see their Figure~9), and 
the results are similar. The same
occurs when we consider a similar hypothetical galaxy cluster at a higher redshift. In this case, the X-ray flux drops
significantly, but so does the tSZ effect as a consequence of the higher beam dilution. Note 
that the value of the tSZ decrement is independent of redshift; however, for a higher redshift the angular size of the cluster is
smaller and then the minimum tSZ decrement convolved with the fixed VSA synthesized beam decreases.
We may therefore conclude that the strongest tSZ effect possible in the VSA map from a galaxy group or a galaxy cluster, 
at any redshift, with a velocity dispersion like the one estimated for CrB-1522+2805 and without detectable X-ray emission in
ROSAT-R6, is $\approx -18~\mu$K, which is 8\% of the total observed decrement in the direction of CrB-H. This 
agrees with the value of  $25^{+21}_{-18}$\% derived from MITO millimetric observations.

\section{Other possible sources of CMB anisotropies toward CrB-H}

As indicated in section~1, in \citet{genova_08} we performed a study based on Monte Carlo simulations, including realizations of
the primordial CMB Gaussian field, residual radio sources and thermal noise, and found the rms introduced by these three 
contributions to be $\approx 52~\mu$K. The observed $-229~\mu$K temperature decrement is therefore a 4.4$\sigma$ deviation. 
When the possible $\approx -18~\mu$K tSZ component is deducted from the CrB-H decrement, it still remains as significant 
4.1$\sigma$ deviation. These results clearly show that the largest tSZ decrement expected from the galaxy group found 
in SDSS-DR7, or from any galaxy cluster without X-ray emission, in spite of being consistent with the MITO result, is unable 
to explain the total decrement in combination with a typical Gaussian primordial CMB anisotropy. Therefore, it is worthwhile to 
consider other sources of secondary CMB anisotropies produced by massive structures, like the Rees-Sciama 
(RS; Rees \& Sciama 1968) effect or the lensing of the CMB.

The RS effect arises when CMB photons traverse time-varying gravitational potentials evolving in the non-linear regime.
When a CMB photon passes through a collapsing (expanding) structure it experiences a net redshift (blueshift) due to the 
different potential wells it faces when falling in and climbing out. The RS effect is then expected to build up CMB 
temperature decrements (increments) in the direction of massive collapsing structures like galaxy clusters (cosmic voids).
Theoretical estimates of the RS effect, carried out by means of the Swiss cheese (SC) model (e.g. 
Martinez-Gonzalez \& Sanz 1990) or the Tolman-Bondi (TB) solution of the Einstein equations (e.g. Saez et al. 1993), give at 
most $\sim -1~\mu$K in the direction of local large-scale structures like the Great Attractor.

\citet{quilis_95} used the TB formalism to estimate the RS effect toward galaxy clusters. For a nearby typical cluster at
$z\approx 0.02$ they estimated a RS decrement of $\approx -30~\mu$K, with an angular size of $\sim 6\degr$. 
An equivalent cluster at a farther distance would generate an effect of similar amplitude but smaller angular size. They remark 
however that this figure is a very soft upper limit, as their assumption of spherical symmetry leads to an excessively rapid 
evolution of the cluster, driven by too fast infalling motions which causes an overestimate of the CMB anisotropy. Conversely, 
in the case of a relaxed galaxy cluster which underwent virialization at $z_{\rm vir}=1$, they obtained a more stringent upper 
limit of $\approx -1~\mu$K.

\citet{dabrowski_99} used the SC model to investigate possible combinations of the RS and SZ effect that could account for 
the intense microwave decrement toward the quasar pair PC~1643+4631 at $z=3.8$ discovered by \citet{jones_97}, without X-ray 
emission \citep{kneissl_98}, and which they suggest could be the result of the SZ effect produced by an intervening galaxy
cluster. They modelled primordial density perturbations of a pressureless field, whose evolution is governed by exact
general-relativistic solutions of the Einstein equations. For a nearby cluster at $z=0.09$ they predicted a RS effect  
with a minimum decrement of $\approx -25~\mu$K, extending up to $\approx 4\degr$, in agreement with 
\citet{quilis_95}. They also note that the spherical free-fall
collapse model they used may not be reliable in the case of such low-redshift clusters which may be virialized. In a more
realistic manner, they applied the same formalism to clusters at $z=1$ with different sizes and densities and obtained a much
larger RS decrement of $\sim -250~\mu$K, with an SZ of $\sim -500~\mu$K. 
It seems therefore that distant galaxy clusters at $z\sim 1$, thanks to their more rapidly-varying gravitational potential,
can build up RS effects of the same order as the SZ effect. For their different cluster models, they found that the
ratio of the RS effect to the SZ effect varies from 0.61 to 0.16 (see their Table~4). Therefore, for the $-40~\mu$K SZ component 
inferred from the MITO observations by \citet{battistelli_06} at CrB-H, we would expect at most a $-24~\mu$K RS decrement.
We have estimated that such clusters would produce X-ray fluxes at the level of the ROSAT-R6 background. 

Massive structures like the CrB-SC and its member galaxy clusters can lense the CMB photons, leading to temperature 
anisotropies. These distortions are usually small and even in the richest galaxy clusters are expected to be of the order of a 
few microkelvins, even though they extend out to angular sizes of a fraction of a degree, and can dominate over the SZ effect in
the outer regions of clusters. \citet{seljak_00} calculated this lensing effect produced in galaxy clusters, and estimated that
its amplitude scales as 10$~\mu$K$\left(\sigma_v/1400~{\rm km~s^{-1}}\right)^2$. According to this, and with the value of the
velocity dispersion of Table~\ref{tab:table_2}, the CrB-1522+2805 group would produce an effect of $\approx 2~\mu$K.

Apart from secondary anisotropies by massive structures, there are other sources of non-Gaussianities on the CMB of primordial
nature. Textures are cosmic defects due to symmetry-breaking phase transitions in the early Universe that can produce negative
and positive fluctuations of the CMB temperature \citep{turok_90}. \citet{cruz_07} argued that this effect could be the cause 
of the $\approx$10$\degr$ `cold spot' detected in WMAP data by \citet{vielva_04}. The predicted number of textures is inversely 
proportional to their angular size squared, and therefore many more are expected with angular sizes similar to the CrB-H decrement. 
However, it is not clear whether textures of sub-degree angular scales really exist, as photon diffusion and other small-scale 
processes could smear them out and current simulations have not enough angular resolution to resolve these angular scales.

\section{Conclusions}
We have analyzed the spatial and redshift galaxy distribution in the SDSS-DR7 spectroscopic catalogue toward the CrB-H 
decrement, a very deep and extended negative feature in the CMB radiation found in a VSA survey of the CrB-SC toward a
position with no known Abell galaxy clusters \citep{genova_05}.
We aimed to explore whether the galaxies trace a filament extended along the line of sight or a previously unnoticed galaxy
cluster at this position, with a significant contribution to the total observed CMB decrement via the tSZ effect. We
found remarkable galaxy overdensities around $z=0.07$, the redshift of the CrB-SC, and $z=0.11$, but no sign of a connection
between these two regions. Therefore, the data does not provide indication of a filamentary structure connecting these 
two regions. In order to asses how significant are the CrB-1522+2858 (the lower-redshift one) and the CrB-1522+2805 
(higher-redshift) overdensities, we have counted the number of galaxies in all the cells of $30\times 30$~arcmin$^2$ and 
$\Delta z=0.01$ within an area of 94.8~deg$^2$ in the region of CrB-SC and between $z=0.05$ and $z=0.12$. We found 
respectively that only 2.6\% and 2.3\% of the cells at redshifts $0.07<z<0.08$ and $0.11<z<0.12$ are denser than 
those corresponding to CrB-1522+2858 and CrB-1522+2805. Furthermore, the majority of these 2.6\% and 2.3\% cells are 
associated with Abell galaxy clusters, indicating that CrB-1522+2858 and CrB-1522+2805 are in fact one of the most 
overdense intercluster regions.

For these two structures, we have analyzed the galaxy number density profiles in both longitudinal and transverse
directions. Whereas CrB-1522+2858 presents a wide redshift distribution, CrB-1522+2805 is steeper and more similar to that of 
the nearby galaxy clusters A2065 and A2069. The radial galaxy number density distributions are flatter than in galaxy clusters 
in both cases, even though CrB-1522+2805 is steeper, with a slightly higher density toward the centre. This indicates 
that CrB-1522+2805 could be somewhat virialized, whereas this is rather implausible for CrB-1522+2858.

We have estimated the baryonic mass of the CrB-1522+2805 structure from its galaxy redshift dispersion. From this 
we calculated the tSZ
effect and the X-ray flux imprinted by this hypothetical structure in the VSA and ROSAT-R6 maps, respectively. Considering
the constraint set by the lack of detectable X-ray emission in ROSAT-R6, we found that the minimum negative tSZ effect
CrB-1522+2805 could produce is $\approx -18~\mu$K, which represents $\approx 8$\% of the VSA decrement. We have 
considered the possibility of a similar galaxy group at higher redshift, but the result is similar. An estimate of 
the SZ effect from the CrB-1522+2858 group is rather implausible, as this structure seems to be unvirialized, which 
makes difficult an estimate of its gas content.
However, an effect of similar amplitude to CrB-1522+2805 cannot be ruled out. 
Our estimated amplitude for the tSZ effect is of 
the same order of the MITO result of 25$^{+21}_{-18}$\% from a spectroscopic analysis of its three frequency channels. 
However, subtracting
the possible $-18~\mu$K tSZ component, the total observed decrement would still remain a significant non-Gaussian deviation 
at the 4.1$\sigma$ level. 

Therefore, the tSZ signal expected from the galaxy groups found in the SDSS-DR7 spectroscopic survey at the position of CrB-H can
only have a minor contribution to the temperature decrement observed in the VSA map, which remains a very significant statistical
deviation from the Gaussian CMB. Other scenarios have then to be considered. One interesting possibility is an 
intrinsic primordial anisotropy in the CMB, but before other mechanisms leading to secondary CMB anisotropies have to be studied. 
The lensing of CMB photons typically produces anisotropies with amplitudes of a few microkelvins in sub-degree angular scales. 
The RS effect in nearby clusters can build up decrements of at most $\sim -20~\mu$K, and in a group of galaxies like 
CrB-1522+2805 it would be much lower. However, more distant and less relaxed clusters, with a faster non-linear variation 
of their gravitational potentials, could give rise to much larger RS signals. Farther rich clusters or groups of clusters 
not identified here could also build strong SZ effects, still compatible with the values allowed by the large error bars of 
the MITO estimate. For this reason, it would be very useful to obtain information of the matter distribution toward this 
position at higher redshifts, by either dedicated X-ray imaging or photometric redshift determinations.

\section*{Acknowledgments}
We are thankful to Marco De Petris and Elia Battistelli for providing comments on this work, to Thomas Reiprich for his 
useful remark about the gas mass fraction in groups of galaxies, and to Juan Betancort for his discussion about the ``$r^2$-test''. 
RGS is funded by the project AYA2007-68058-C03-01 of the Spanish Ministry of Science and Innovation. JARM is a Ram\'on y Cajal 
Fellow of the Spanish Ministry of Science and Innovation. 

The SDSS is managed by the Astrophysical Research Consortium for the Participating Institutions. The Participating Institutions 
are the American Museum of Natural History, Astrophysical Institute Potsdam, University of Basel, University of Cambridge, 
Case Western Reserve University, University of Chicago, Drexel University, Fermilab, the Institute for Advanced Study, the Japan Participation Group, Johns Hopkins University, the Joint Institute for Nuclear Astrophysics, the Kavli Institute for Particle Astrophysics and Cosmology, the Korean Scientist Group, the Chinese Academy of Sciences (LAMOST), Los Alamos National Laboratory, 
the Max-Planck-Institute for Astronomy (MPIA), the Max-Planck-Institute for Astrophysics (MPA), New Mexico State University, 
Ohio State University, University of Pittsburgh, University of Portsmouth, Princeton University, the United States Naval 
Observatory, and the University of Washington.

\end{document}